# Local Enhancement of Lipid Membrane Permeability Induced by Irradiated Gold Nanoparticles

*Andrea Torchi[1], Federica Simonelli[1], Riccardo Ferrando[2] and Giulia Rossi[1]\**

[1] *Physics Department, University of Genoa, via Dodecaneso 33, 16146 Genoa, Italy*

[2] *Chemistry Department, University of Genoa, via Dodecaneso 31, 16146 Genoa, Italy*

KEYWORDS

lipid membranes, gold nanoparticles, photothermal applications, photoporation, molecular dynamics

ABSTRACT

Photothermal therapies are based on the optical excitation of plasmonic nanoparticles in the biological environment. The effects of the irradiation on the biological medium depend critically on the heat transfer process at the nanoparticle interface, on the temperature reached by the tissues as well as on the spatial extent of temperature gradients. Unfortunately, both the temperature and its biological effects are difficult to be probed experimentally at the molecular scale. Here, we approach this problem using non-equilibrium molecular dynamics simulations. We focus on photoporation, a photothermal application based on the irradiation of gold nanoparticles by single, short-duration laser pulses. The nanoparticles, stably bound to cell membranes, convert the radiation into heat, inducing transient changes of membrane permeability. We make a quantitative prediction of the temperature gradient around the nanoparticle upon irradiation by typical experimental laser fluences. Water permeability is locally enhanced around the NP, in an annular region that extends only a few nm far from the nanoparticle interface. We correlate the local enhancement of permeability at the NP-lipid interface to the temperature inhomogeneities of the membrane and to the consequent availability of free volume pockets within the membrane core.

Metal nanoparticles (NP) can absorb light in the visible and near infrared (NIR) window and convert it into heat. As NIR radiation can penetrate biological tissues for several centimeters, the ability to deliver metal NPs to a target unhealthy tissue allows for the localized release of heat in the region to be treated. Photothermal applications are currently the object of intense research efforts,[1-3] aimed at the optimization of the NP composition and optical response,[4,5] as well as at the minimization of their toxic effects.[4] Gold plasmonic nanoparticles, which show good biocompatibility, have already entered clinical trials for the treatment of cancer.[6,7] The capability of Au nanoparticles to convert light into heat can be exploited also for drug delivery applications. Polymer capsules or liposomes loaded with nanoparticles[8,9] have been shown to release their cargo upon irradiation.[10-12] The same physical principle holds in photoporation applications,[13] where irradiated Au NPs allow for the transfer of macromolecules into the cell, for example for transfection.[14]

The effectiveness of the photothermal therapy or delivery, as well as its possible adverse effects, depend on a number of physico-chemical characteristics of the nanoparticle, such as its composition, its size and shape and the chemical nature of its protecting organic monolayer. The relation between the NP size, shape and optical properties has been thoroughly investigated at experimental and theoretical level.[15-17] It is less clear how heat is then transferred to the surrounding biological medium. The measurement of the local temperature in the cellular environment is extremely challenging from the experimental point of view,[18,19] especially when one aims at measuring temperature gradients with sub-nanometer resolution around a heated NP.[20]

Here we propose a molecular dynamics (MD) study of the localized heating of a model lipid bilayer in presence of an irradiated Au nanoparticle. The damage of the lipid bilayer by localized hyperthermia is indeed a key factor in all photothermal applications. In photothermal therapies, and tumor treatments in particular, the increased permeability of cell membranes either directly affects the viability of cancer cells, or allows for the uptake of drugs or further nanoparticles. In liposome-based drug delivery approaches, the membrane disruption or phase change[21-24] allows for the release of the cargo drug. In photoporation assays, transfection is made possible by the transient poration of the cell membrane,[14] with little or no cytotoxic effects.

The relevant length and time scales for photoporation applications can be handled by state of the art molecular simulations. The damage induced on the membrane is transient and thus spatially limited to the nm scale, while typical laser impulses have a duration of a few ns.[25] Several challenges, though, need to be faced. The first is to develop atomistic models that reliably describe the vibrational properties of the Au NP and, at the same time, allow for computational efficiency. Here we develop an atomistic model of a Au NP that reproduces realistically its harmonic vibrational spectrum and is compatible with a popular atomistic force field used to describe the biological environment. We validate our model through the calculation of the thermal conductance at the NP interface, which turns out to be consistent with the available experimental data on thiolated Au surfaces.[26] Another critical issue is the simulation of the constant energy flow that is transmitted during single laser impulses to the NP. We thus set up an MD protocol to handle typical laser fluences used in photoporation experiments. Our simulations predict that the lipid membrane can reach temperatures of about 370 K around the NP, with important effects on membrane structure and permeability. We find that water permeability is locally enhanced at the NP-lipid interface. This enhancement of permeability is correlated to the temperature inhomogeneities of the membrane, and to the consequent availability of free volume pockets within the membrane core.

By offering a molecular-level interpretation of the damage induced to the membrane by localized heating, we can understand which properties of the NP, and of the NP-membrane complex, can be tuned to achieve a quantitative control on the transient modification of membrane permeability.

RESULTS AND DISCUSSION

**The configurations of the NP-membrane complex and the Au NP model**
We consider a flat lipid membrane, composed by 1-palmitoyl-2-oleoyl-sn-glycero-3-phosphicholine (POPC) lipids, interacting with a single Au NP with a diameter of 2 nm.

The size of the NP is much smaller than those typically used in photoporation experiments for reasons of computational feasibility. The Au core, whose structure has been determined *via ab initio* calculations,[27] is covalently functionalized *via* Au-S bonds by a mixture of 30 hydrophilic (negatively charged 11-mercapto-undecane-carboxylate, MUC) and 30 hydrophobic (octanethiol, OT) ligands, as shown in Figure S1. This functionalization assures a stable binding of the NPs to PC bilayers.[28,29] The NP-membrane complex is modeled with united-atom resolution in explicit water, as described in the Methods section. In our previous works[30,31] we have identified two conformations of the NP-membrane complex that are stable, at physiological conditions, over timescales of at least tens of microseconds. In the first case (Figure 1a) the NP is only partially embedded in the membrane. The NP center of mass is located in the lipid head region of one bilayer leaflet, and the complex is stabilized by the hydrophobic contact between the hydrophobic moieties of the ligands and the lipid tails. We will refer to this state as to the hydrophobic contact (HC) state. In the second case (Figure 1b), the NP is fully immersed in the lipid membrane, with all its hydrophobic ligands facing the membrane core and the hydrophilic ligands interacting with the headgroups of both leaflets. We will refer to this state as to the anchored (ANCH) state.

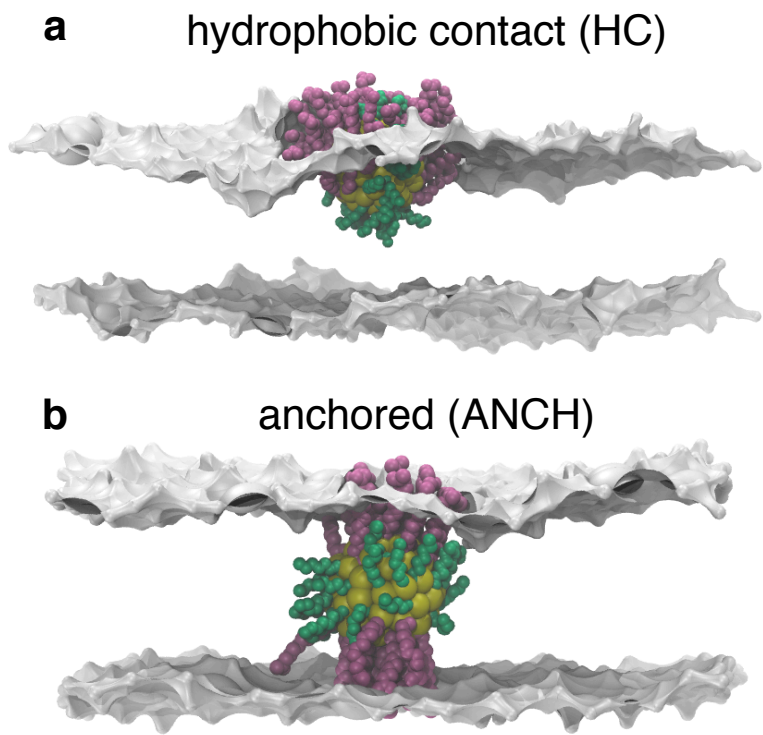

**Figure 1** – Gold and sulfur atoms in yellow, the anionic ligands in pink, the hydrophobic ligands in green. Lipid heads are shown in grey, lipid tails and water not shown. (a) A NP in the hydrophobic contact configuration, interacting with the upper membrane leaflet (b) A NP in the anchored configuration, in which its anionic ligands interact with both membrane leaflets.

Within a classical MD approach, the transfer of heat from the irradiated NP to the ligands, solvents and membrane can include only the phonon contribution, which is anyway the most relevant one. While it is computationally convenient to rely on pair potentials to describe Au-Au interactions, the Lennard-Jones potential that is used by the OPLS united atom force field suffers from intrinsic limitations at describing the many body character of metal bonding.[32] A better description of Au-Au interactions is provided by the many-body potential energy model derived from the second moment approximation to the tight binding scheme.[33] This potential takes into account the bond-order/bond-length correlation that is typical of metals:[34] low-coordinated atoms (surface atoms) are bound by stronger and shorter bonds than highly coordinated (volume) atoms. We thus consider as a target property the harmonic vibrational spectrum of the Au NP given by this many-body model, and parameterize an elastic network in such a way as to reproduce the spectrum as well as possible. The optimized Au NP elastic network uses two different elastic constants, a larger one for the surface Au atoms (32500 kJ mol⁻¹ nm⁻¹) and a smaller one for the atoms of the core (11000 kJ mol⁻¹ nm⁻¹), effectively capturing the difference between surface and volume bonds that is intrinsic to the many-body approach. In Figure 2 we show the overlap between the harmonic vibrational spectra of the NP as predicted by the many-body potential and by the optimized elastic network.

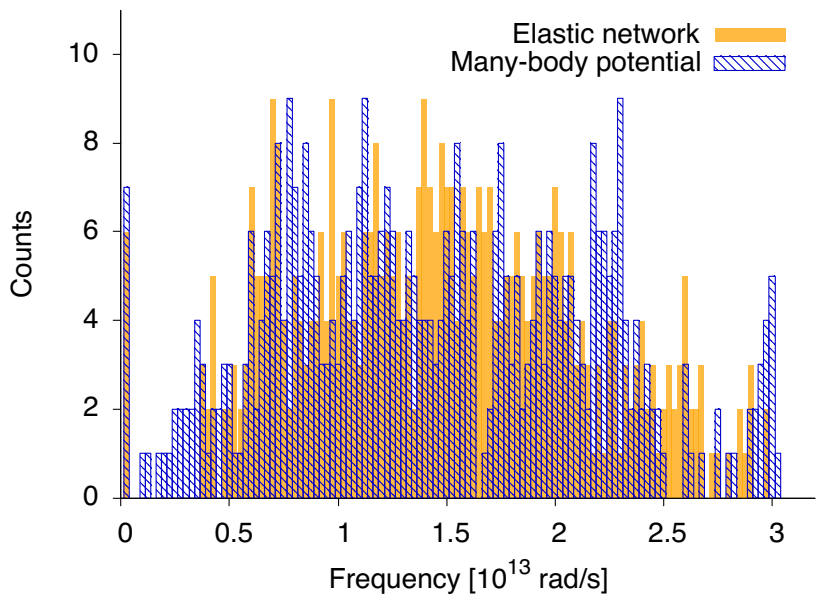

**Figure 2 – Comparison between the vibrational spectra of the Au₁₄₄ NP predicted by the many-body potential and by our elastic network model.**

### Setup no.1: NP thermal conductance

Our first simulation set-up is aimed at verifying that the thermal conductance of the Au NP, as predicted by our model, is consistent with the available experimental and theoretical data. After equilibrating the NP-membrane complex in the HC configuration at 310 K and atmospheric pressure, the entire system is decoupled from the thermostat and only the NP Au core is thermalized at a temperature of 400 K,[35] thus mimicking the selective absorbance of an incident radiation. This target Au core temperature is low enough not to lead to the melting of the ligand-protected NP,[36] and at the same time allows us to monitor the progressive heating of the surrounding environment over the typical time scales of the laser pulses used in photothermal applications. Figure 3a shows how the ligands, water and lipids gradually approach the Au core temperature during a 20 ns run. The transfer of heat from the Au core to the water and lipid environment is mediated by the layer of organic ligands. The thermal conductivity, $k$, at the interface between a spherical NP of radius $r$ and a surrounding fluid phase is defined by the law of thermal conduction:

$$\frac{\partial Q}{\partial t} = -k \oiint \nabla T(\vec{r}) d\vec{S} \tag{1}$$

where $\partial Q/\partial t$ is the heat transfer rate through the closed surface $S$, and $T$ is the temperature on the surface. If the temperature inhomogeneities (water vs. lipids) in the system surrounding the NP core can be considered negligible with respect to the temperature difference between the core and the surroundings, equation (1) can be simplified into:

$$\frac{\partial Q}{\partial t} = -k'(T_{water} - T_{Au}) \tag{2}$$

The quantity $k'/S$ is the heat transfer coefficient. We have verified that in our simulations the relation between $\partial Q/\partial t$ and $(T_{water} - T_{Au})$ is indeed linear to a good approximation (see Figure S2). By setting the radius of the interface $r$ at 1 nm we derive a heat transfer coefficient of 97 MW m$^{-2}$ K$^{-1}$. The presence of the organic ligand layer affects the interfacial thermal conductance, which is indeed lower than that calculated by MD simulations at the NP/water interface for bare Au NP.[37] Our transfer coefficient is in line with previous computational estimates for self-assembled monolayers.[38,39] It falls well within the range of those reported experimentally for hydrophilic (alkane layers with polar terminal groups, 150 MW m$^{-2}$ K$^{-1}$) and hydrophobic (alkane layers with methyl terminals, 50 MW m$^{-2}$ K$^{-1}$) interfaces, coherently with the heterogeneous nature of the ligand layer surrounding the Au core. We are on the low side of the 100-300 MW m$^{-2}$ K$^{-1}$ range experimentally measured for AuPd NPs stabilized by organic ligands[40,41] of different hydrophobicity.

**Setup no.2: simulations with a constant energy inflow**

In setup no.1 the energy flow from the thermostat to the system is not constant over time: as the system temperature gets closer to the Au thermostat target temperature, the energy flux decreases. In typical photoporation experiments, the NPs are irradiated with single laser pulses, during which the flux can be considered constant. In order to keep the flux of energy from the thermostat to the NP constant, we progressively (every 0.2 ns) increase the target Au temperature, $T_{Au}$, according to:

$$T_{Au} = \frac{P}{k'} + T_{water}$$

$P$ being the laser power. Assuming a 100% conversion efficiency,[16] a pulse duration $\Delta t = 10$ ns and a laser fluence $F = 550$ J/m² we can set $P = \frac{F \mathcal{N}_A A}{\Delta t}$ at 103 kJ mol⁻¹ ps⁻¹. The value of the laser fluence is consistent with those used in photoporation experiments and should lead to membrane transient permeation without causing the formation of vapor nanobubbles[14] or the NP melting.[42,43]

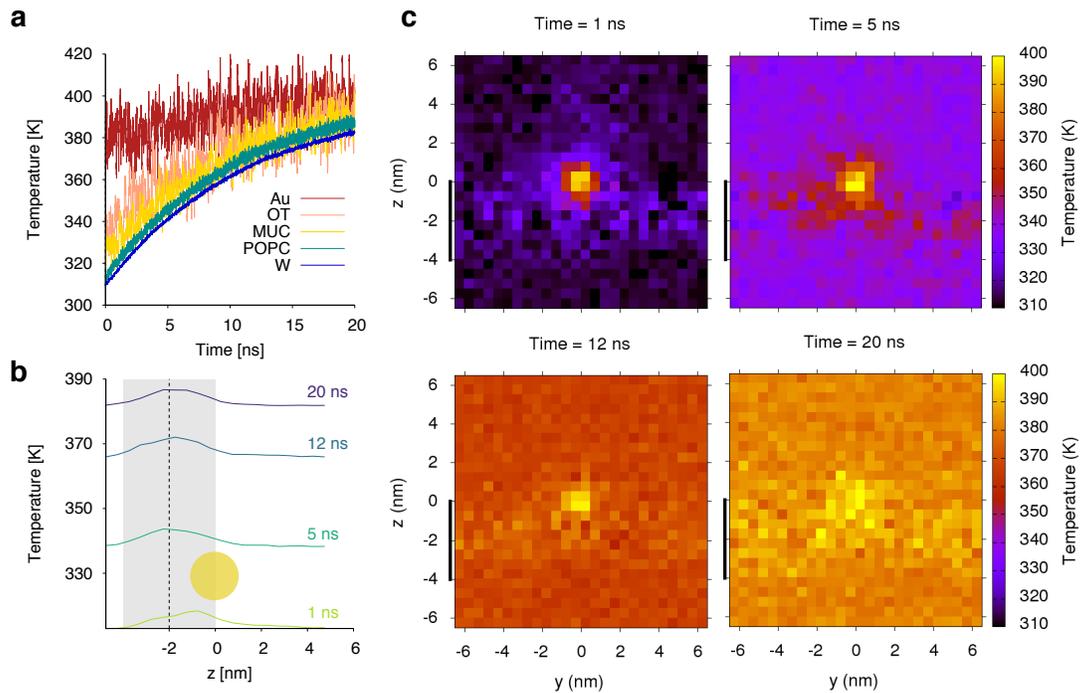

**Figure 3** – **a.** Temperature (averaged over all molecules) increase of the different system components during the run with setup no.1. Au refer to the NP atoms, OT and MUC to the hydrophobic and anionic ligands, respectively, POPC refer to lipids and W to water; **b.** Temperature profile along the z direction. Here, the yellow circle indicates the position of the NP, while the shaded grey area indicates the membrane position; **c.** Temperature maps on the yz plane that crosses the NP center of mass. The black bar indicates the position of the membrane.

Constant energy inflow simulations can be performed in two different ensembles. In the first ensemble we consider, the volume of the simulation box is kept constant and, as heat is transferred from the thermostat to the Au NP, both the pressure and the temperature in the box increase. We will refer to the first ensemble as to the $NVT_{Au}$ ensemble, where $T_{Au}$ means that Au is the only part of the system that is subject to the action of the thermostat. In the second ensemble, instead, the pressure in the simulation box is kept constant at 1 bar, the barostat acts as an energy sink and the temperature of the box increases at a slower rate. We will refer to the second ensemble as to the $NPT_{Au}$ ensemble. On the one hand, an increase of pressure around the NP is not unrealistic, and indeed pressures up to 30 MPa are theoretically predicted at the onset of irradiation-induced nanobubble formation.[44] On the other hand, in our $NVT_{Au}$ ensemble the pressure reaches much larger values (up to 100MPa) after a 10 ns irradiation. We thus resolve to perform heating simulations in the $NPT_{Au}$ ensemble, bearing in mind that this setup could imply an underestimation of the temperature increase of the water and membrane phases. As shown in Figure S3, such an underestimation should not exceed 20 K. Figure 4 shows the temperature increase of the system components as a function of time in a $NPT_{Au}$ simulation.

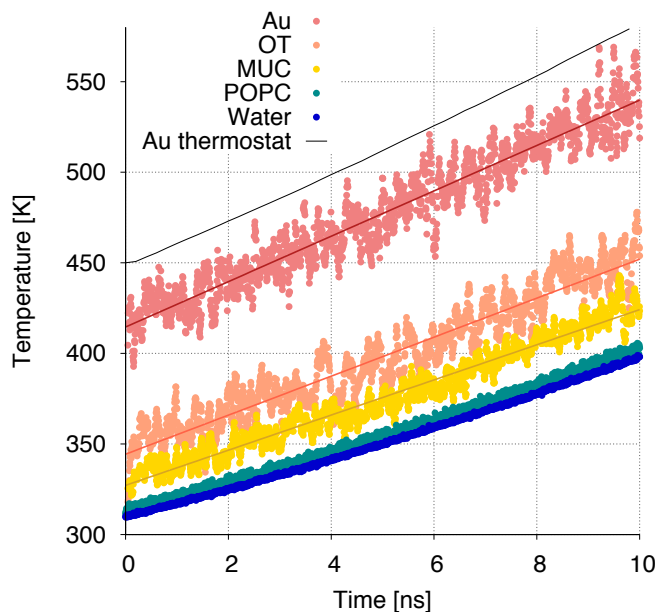

**Figure 4 – The temperature increase of the Au NP (red), OT ligands (orange), MUC ligands (yellow), POPC (green) and water (blue) using the $NPT_{Au}$ ensemble and a laser fluence F = 544 J/m². The black line indicates the Au thermostat temperature. Straight lines are linear fits to the data.**

According to our model and with the aforementioned settings, a single laser pulse with a duration in the 7-10 ns range[43] should indeed be sufficient to bring water close to its liquid to gas transition temperature (assuming no pressure increase) and the monolayer-protected NP around its melting temperature, which sets at 500 K.[42]

**Effects of the irradiated NP on the membrane permeability to water**

In photoporation applications, several NPs interact with each cell membrane and the laser scanning speed can be regulated so as to irradiate each NP with a single laser pulse only.[43] Our scope is to investigate the effects that the NP and the temperature increase can have on membrane permeability, and thus we resolve to exploit the ergodic hypothesis and perform longer simulations (100 ns) in the NP{$T_i$} ensemble, using different thermostats to couple each system component ($i$ = NP, OT, MUC, POPC, water) to the temperature predicted by our heating setup (Figure 4) at different times. Table 1 summarizes the NPT simulations performed and analyzed in the following.

Table 1. Simulations in the NP{$T_i$} ensemble.

| $T_{Au}$ [K] | $T_{POPC}$ [K] | System configuration |
|---|---|---|
| 460 | 320 | HC |
| 480 | 330 | HC/ANCH/NO NP |
| 498 | 340 | HC |
| 506 | 345 | HC/ANCH/NO NP |
| 514 | 350 | HC |
| 525 | 355 | HC/ANCH/NO NP |
| 531 | 360 | HC |
| 539 | 365 | HC/ANCH/NO NP |
| 310 | 310 | HC/ANCH/NO NP |
| 365 | 365 | HC/ANCH/NO NP |

The table reports the temperature of the POPC and NP thermostats (as shown by the green and black lines in Figure 4, respectively). The temperature of water, as shown in Figure 4, differs from that of POPC by a few degrees only. All simulations are 100 ns long. HC and ANCH are two different configurations of the NP-membrane complex (Figure 1). NO NP means that a replica of the simulation has been performed for a system composed of POPC and water only. The last two simulations have been used to characterize the structural features of the membrane, at the lowest and highest POPC temperature, in absence of NP-membrane temperature gradient.

We then perform a direct calculation of water permeation as a function of the $T_{POPC}$ temperature. Membrane permeation events were recorded by an *ad hoc* analysis tool, as described in the Methods section. Single water permeation events span timescales of several hundred picoseconds[45,46] at the lowest temperature of 310 K, where in a few cases water molecules wander within the membrane core for more than 1 ns. At the highest temperature of 365 K, the distribution of the permeation times is narrower and peaked at 150 ps, as shown in Figure S4. Figure S5a shows the number of permeation events per unit area as a function of the POPC temperature. By fitting the data to an Arrhenius behavior, we get an activation energy, $E_a$, of about 48 kJ/mol for both the HC and the ANCH configuration, in agreement with previous atomistic simulations[47] and within the broad range of experimental estimates.[48,49]

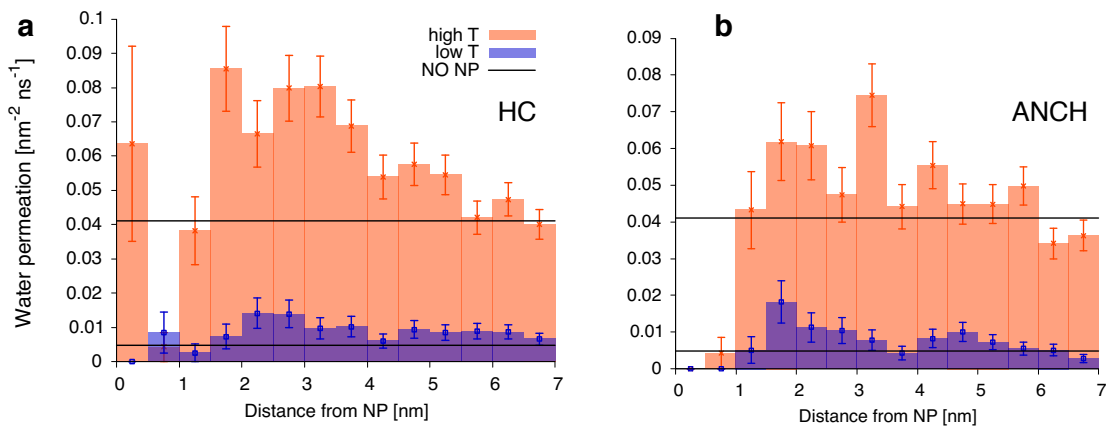

**Figure 5.** Water permeation rate as a function of the distance from the NP center of mass for the HC (a) and the ANCH (b) configuration. Low T data refer to $T_{POPC}$ = 330 K, high T data to $T_{POPC}$ = 365 K

**Temperature inhomogeneities and NP-induced local disorder concur at increasing membrane permeability**

We now look in more depth at the permeation data. Each permeation event is characterized by a point of entrance into the membrane and a point of exit from the membrane (details in the Methods section). Figure 5a and 5b show the permeation rate as a function of the distance, $d_p$, between the entry point of the water molecule and the center of mass of the NP, in the *xy* membrane plane. The data are referred to $T_{POPC}$ = 330 K and $T_{POPC}$ = 365 K. Permeation events are not uniformly distributed in the membrane plane, but are more probable in a region between the NP-membrane interface, at $d_p$ = 1.5 nm, and $d_p$

= 4 nm. This trend is especially visible at high temperature. In Figure S5b and S5c we plot the number of permeation events recorded in two regions, close to the NP ($0 < d_p < 4$ nm) and far from the NP ($5 < d_p < 7$ nm), as a function of temperature. The fit to the Arrhenius behavior provides only slightly different barriers for the permeation processes in the two regions.

We can rationalize the local enhancement of permeability around the NP in terms of the structural and thermal inhomogeneities induced by the irradiated NP in the membrane. In our NP{$T_i$} set up, each system component is thermalized at a different temperature. Nevertheless, the $T_{POPC}$ reported in Table 1 are only average temperatures of the lipid membrane, and indeed we observe a non-uniform temperature distribution around the NP, as shown in Figure 6.

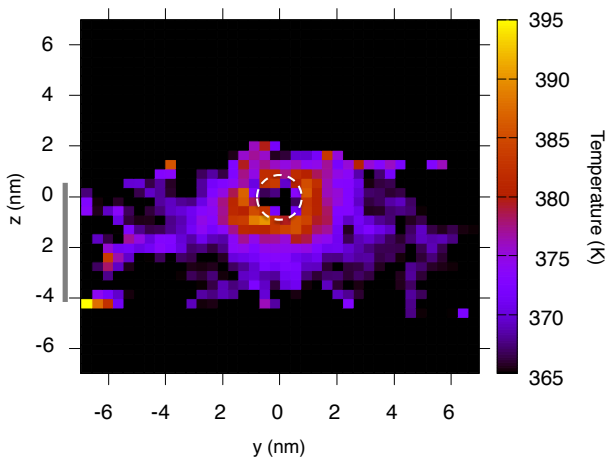

**Figure 6. The time-averaged temperature of lipids in the NP{$T_i$} ensemble, with $T_{POPC}$ = 365 K and the NP in the HC configuration. The white circle indicates the position of the NP Au core.**

Figure 7 shows the analysis of the area per lipid calculated for the HC (panels a, b) and ANCH (panels c and d) configurations, as a function of the distance from the NP center of mass. In the HC configuration, the NP sensibly affects the area per lipid in a region that extends roughly from 2 to 4 nm from the NP. The area per lipid is only slightly reduced in the distal leaflet, while it undergoes large oscillations in the leaflet hosting the NP. The effect is present both at the lowest (310 K) and highest (365 K) $T_{POPC}$ we considered. The area per lipid is instead very little affected by the NP in the ANCH configuration, where it increases only at a distance of 2 nm from the NP, where the lipids overlap with the NP ligands. The area per lipid around the NP is not more sensitive to temperature increases than it is far from the NP. The fact that the area per lipid changes very little or even decreases (in the HC configuration) in the region where we register the largest number of permeation events lead us to conclude that the area per lipid is not responsible for the NP-induced local enhancement of permeation.

If the reason for the local enhancement of permeation is not the availability of free surface area, we proceed to the quantification of the free volume within the membrane. The two NP configurations affect the local free volume to a similar extent (Figure 7, panels e and f, details on the calculation of the free volume can be found in the Methods section). The free volume perturbation is limited to a rather confined cylindrical shell around the NP, whose thickness is about 1.5 nm. The availability of free volume in the membrane has been shown to correlate well,[47,50,51] though only qualitatively,[52] with the diffusion rates of different membrane penetrants, and thus it is expected to affect water permeability. In the NP-membrane complex, the shape of the free volume profile seems to be particularly sensitive to temperature increases around the NP.

**Conclusions**

In this paper we have investigated, by classical Molecular Dynamics with united atom resolution, the process by which a surface plasmon resonant Au NP releases heat to a lipid bilayer, in presence of a short pulse laser stimulus. The Au NP, covalently functionalized by a mixture of hydrophilic and hydrophobic ligands, can be stably bound to the membrane surface or to the membrane core. We predict a heat transfer coefficient of 97 MW $m^{-2}$ $K^{-1}$ at the NP interface. For typical laser pulses of 10 ns and laser fluences of 550 J/$m^2$, the NP heats up to about 500 K while the lipid bilayer can reach average temperatures of about 370 K. The lipid bilayer may reach temperatures around 385 K in proximity of the NP surface. The temperature difference between the NP and the surrounding bilayer is consistent with the available experimental estimates.[20] As the temperature increases, the membrane gets more permeable to water, an effect that is locally enhanced around the NP. Such an enhancement, we find, is due to spatial temperature inhomogeneities. The increase of the local temperature of lipids significantly changes the free volume at the NP-membrane interface, a structural feature that is indeed correlated with increased permeability.

We anticipate that our approach can be developed in several directions. The size of the NPs used in photoporation experiments is typically larger than that used in this work, as larger NPs are known to have more peaked SPR spectra. Larger NPs can be less prone to interact passively with cell membranes: typically, NPs with a diameter larger than 10 nm do not spontaneously enter the membrane core. They can, though, stably interact with the membrane surface and cause the formation of membrane pores.[14] Our approach could be directly applied to this latter case, provided the NP size does not grow up to the point to be computationally unmanageable at atomistic level.

Our approach to the development of a metal core model that has the correct vibrational spectrum and, at the same time, is computationally efficient, could be directly applied to different core materials. Other compounds, such as copper sulfide,[4] could indeed be evaluated to obtain better resonances in the NIR spectrum, which would in turn guarantee a deeper penetration in biological tissues. An important role may be played by the NP ligands, as well, as they may tune the effectiveness of heat transfer at the nanoparticle surface. These effects are still largely unexplored, both at the experimental and computational level.

Another interesting route of development will concern the modeling of lipid heterogeneity and of other membrane inclusions, as well as the presence of clusters of NPs interacting with the bilayer.

In conclusion, there is a wide range of photothermal applications that could take advantage of the molecular resolution offered by molecular dynamics. Simulations can be used to investigate the effects of localized heating on other biological environments, and they could as well be exploited to support the design of drug delivery systems, such as light-responsive liposomes or polymer capsules.

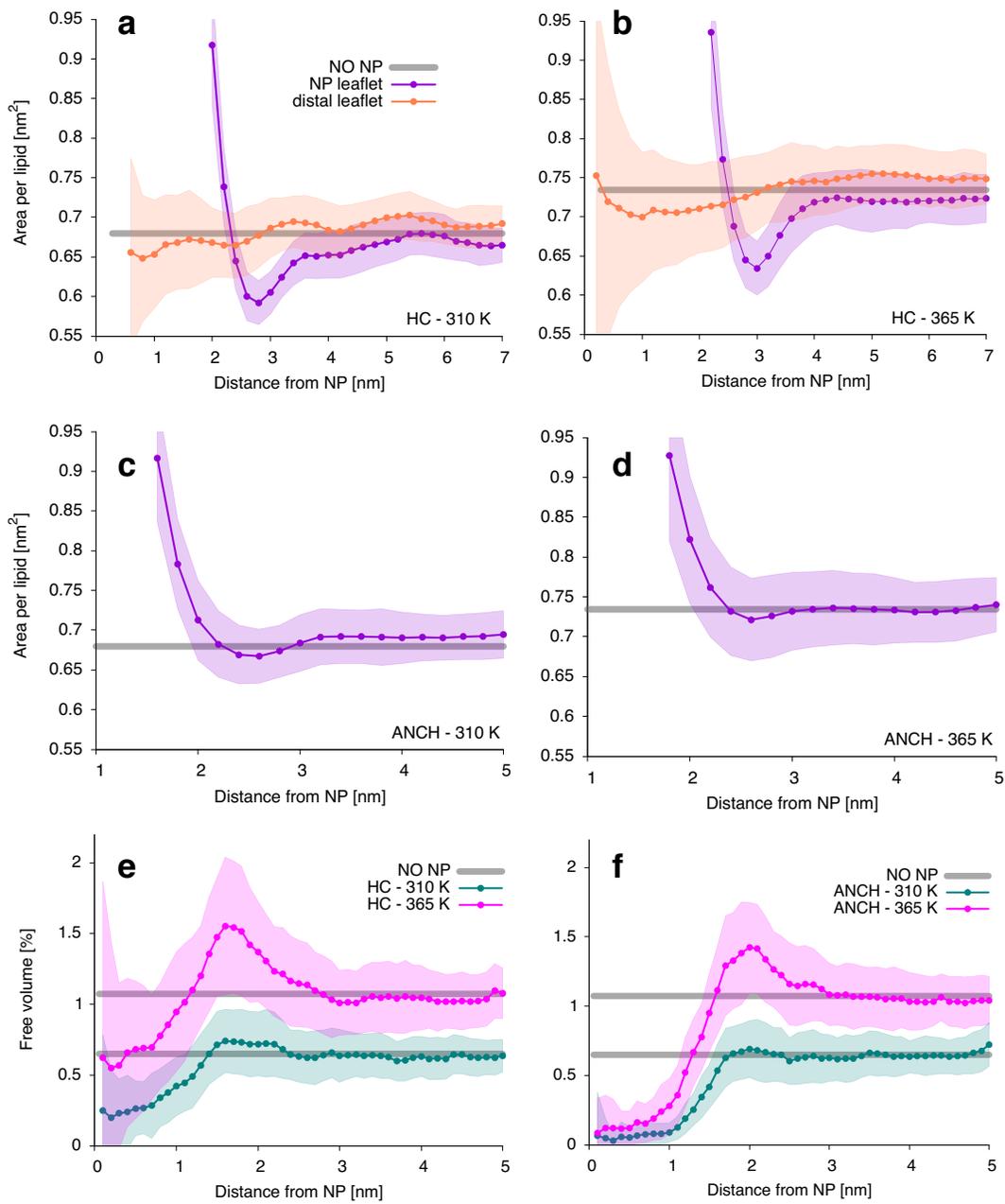

**Figure 7. NP-related structural inhomogeneities of the membrane at 310 K and 365 K.**

## Methods

*Au NP model*. The core of the NP, as derived by Lopez-Acevedo[27] *et al*., is made of 144 Au atoms with icosahedral symmetry and 60 S atoms bound to the gold core. Both Au and S atoms are connected through an elastic network, as described in the main text. A total of 60 ligands are bound to the NP core *via* Au-S bonds. The ligands are 30 hydrophobic OT and 30 negatively charged MUC. The chemical structure of the ligands is shown in Fig. S1. MUC and OT interactions are parameterized with the united-atom OPLS force field.[53]

*The NP-membrane complex*. All simulations involving the NP in the hydrophobic contact configuration were performed with a membrane of 706 POPC lipids, corresponding to a simulation box of approximately 15.5x15.5x11.0 nm$^3$ at 310 K. All simulations involving the NP in the anchored configuration were performed with a membrane of 460 POPC lipids, corresponding to a simulation box of approximately 12.6x12.6x10.1 nm$^3$ at 310 K. NP ligands were modeled with the united atom OPLS force field.[54] Lipid interactions were modeled with Berger parameters.[55] We used SPC water. All simulations were done at physiological (150 mM) salt concentration. Long-range electrostatics were included *via* particle-mesh-Ewald summation, with a Fourier grid spacing of 0.12 nm.

*MD simulation set up*. All simulations of the heating process (setup no.1 and no.2 in the main text) were performed by applying a thermostat to the Au NP core only, thus leaving the rest of the system (ligands, lipids, water and ions) not thermalized. In these simulations, the flexible version of SPC water was used. We tuned the MD parameters in such a way as to obtain good energy conservation (a temperature drift of less than 0.003 K/ns) in a box of water at 370 K. We used the double precision version of Gromacs 5.1.4 and a velocity Verlet integrator. The timestep was set to 1 fs; Au was thermalized with the Bussi-Parrinello thermostat,[56] with a characteristic time of 0.5 ps (larger than the fastest vibrations of the Au core); neighbor-list updating was performed with the Verlet algorithm, with a neighbor list radius (rlist in Gromacs) of 1.1 nm. In the NPT$_{Au}$ ensemble, the Berendsen barostat[57] was used, with a time constant of 1.0 ps and semi-isotropic coupling (*x* and *y* directions coupled separately from the *z* direction). All simulations performed in the NP{T$_i$} ensemble were performed with the single precision version of Gromacs 5.1.4 and the rigid version of SPC water, the other parameters unchanged.

*2D lipid temperature map.* The temperature map in Figure 6 shows the time-averaged temperature of lipids as obtained from the 100 ns run in the NP{$T_i$} ensemble, with $T_{\{POPC\}}$ = 365 K and $T_{\{Au\}}$ = 539 K. The NP was in the HC configuration. We considered only the lipids inside a *yz* slab that contained the center of mass of the NP. The slab thickness, along the *x* direction, was 2 nm, corresponding to the NP core diameter. The trajectory was sampled every 0.1 ns along the 100 ns run, and average lipid temperatures are shown on the color map only for the *xy* grid points for which more than 50 samples were collected.

*Area per lipid.* The nonlocal area per lipid, shown in Figure S6, was always calculated as the ratio between the average area of the simulation box and the number of lipids in each leaflet. We thus remark that in the simulations comprising the NP the area per lipid is necessarily larger than in the unperturbed membrane. The local area per lipid (Figure 7) was calculated as the ratio between the area of concentric circular xy shells centered on the NP center of mass and the number of lipids present in the shell. The number of lipids in each shell was time averaged along the 100 ns run in the NP{$T_i$} ensemble.

*Tracking of the water permeation events.* To track water permeation events, we saved the system configurations every 1 ps, corresponding to a water mean displacement of 0.1 nm at 365 K. A permeation event is recorded any time a water molecule enters the membrane from the top (or bottom) leaflet and exits the membrane from the bottom (or top) leaflet. Water molecules that enter and exit the membrane from the same leaflet do not contribute to the count of permeation events. A water molecule is considered to have entered the membrane when its z coordinate is $z_{up} > z > z_{down}$, where $z_{up}$ and $z_{down}$ correspond to the average positions of the phosphate groups of the two membrane leaflets. The algorithm takes into account membrane oscillations by calculating these membrane boundaries locally, on a *xy* grid with bins of 3 nm. Boundaries are updated every 1 ps.

*Free volume calculations.* For the calculation of the free volume we first identified a 3D slab comprising the membrane along the whole simulation. A slab with a thickness of 7 nm was typically sufficient to take into account membrane oscillations. Then, the slab was divided into a 3D grid with bins of 0.1 nm. Each grid cube contributed to the free volume if its minimum distance from any atom in the system was larger than 0.3 nm. The free volume reported in Figure 7 is calculated as the ratio between the number of free grid cubes and the total number of grid cubes in the slab.

**Corresponding author**

Giulia Rossi, email: rossig@fisica.unige.it

**Author contributions**

R. F. and G. R. conceived the idea, planned the simulations and supervised the work. F. S. developed and tested the Au nanoparticle model. A. T. performed most of the simulations, developed the analysis software and, together with G. R., analyzed the data. G. R. wrote the paper with the collaboration of all authors.

ASSOCIATED CONTENT

**Supporting Information.** The Supporting Information Available online file includes a) Figure S1 showing the atomistic details of the Au NP model; b) Figure S2 and S3 with details concerning our heating simulation protocols; c) Figures S4, and S5 showing the time distribution of water permeation events and an Arrhenius plot of the water permeation events as a function of temperature; d) Figure S6 showing the variations of membrane thickness and area per lipid as a function of temperature.

ACKNOWLEGEMENT

Giulia Rossi acknowledges funding from the ERC Starting Grant BioMNP – 677513. Part of the calculations was performed at CINECA under the HP10CRSL8N grant. The authors thank Sebastian Salassi for fruitful discussions.

The authors declare no competing financial interests.

# Local enhancement of lipid membrane permeability induced by irradiated gold nanoparticles

*Andrea Torchi[1], Federica Simonelli[1], Riccardo Ferrando[2] and Giulia Rossi[1]\**

[1] *Physics Department, University of Genoa, via Dodecaneso 33, 16146 Genoa, Italy*
[2] *Chemistry Department, University of Genoa, via Dodecaneso 31, 16146 Genoa, Italy*

**Supplementary material**

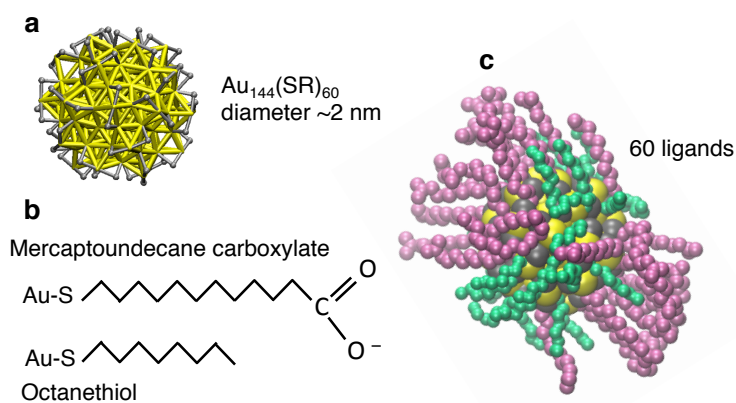

**Figure S1 – a.** The elastic network chosen to model the Au NP core. Yellow: Au atoms, grey: S atoms; **b.** The ligands covalently bound to the NP surface via Au-S bonds; **c.** the complete NP+ligands structure. The hydrophobic ligands (green) are placed on an equatorial stripe around the NP, to assure a stable binding to the membrane[1,2]. The hydrophilic ligands are shown in pink.

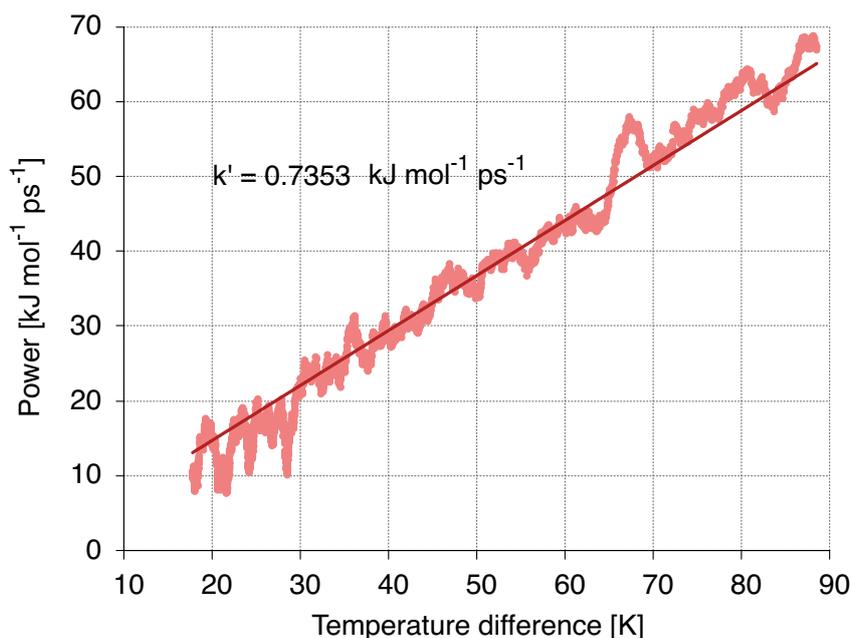

**Figure S2 –** The time derivative of total energy vs. the temperature difference between the Au thermostat temperature in setup no.1, $T_{Au}$ = 400 K, and the average water temperature.

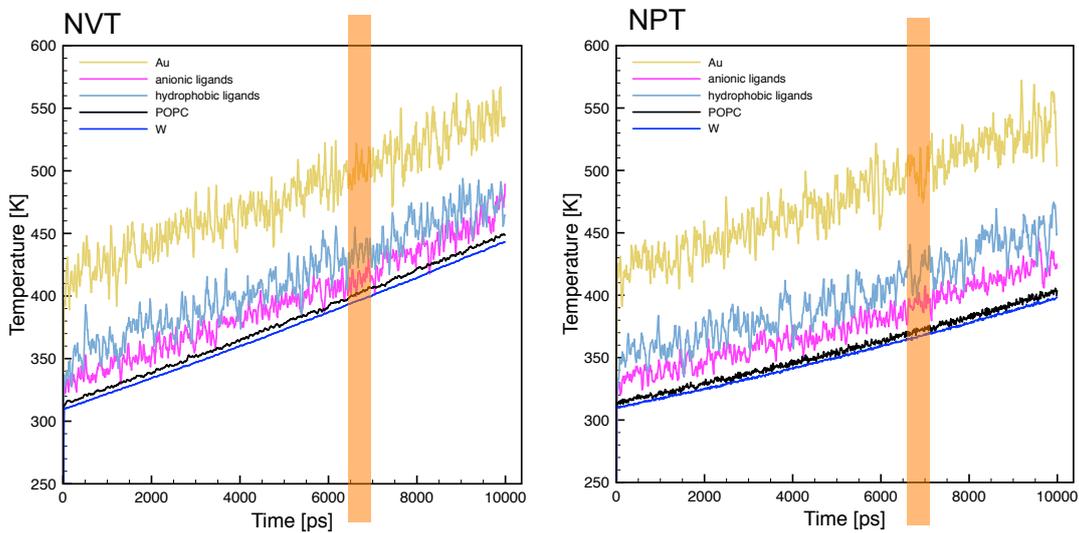

**Figure S3** – A comparison between the constant energy inflow (setup no.2) simulations, in the NV{$T_{Au}$} (left) and NP{$T_{Au}$} (right) ensembles. In the NP{$T_{Au}$}, the barostat acts as an energy sink and the water and membrane temperature increase at a slower pace. The orange bar indicates the time at which the NP reaches its melting temperature. In those conditions, the NP{$T_{Au}$} simulations predicts that the POPC has reached a temperature that is about 20 K below that reached in the NV{$T_{Au}$} ensemble.

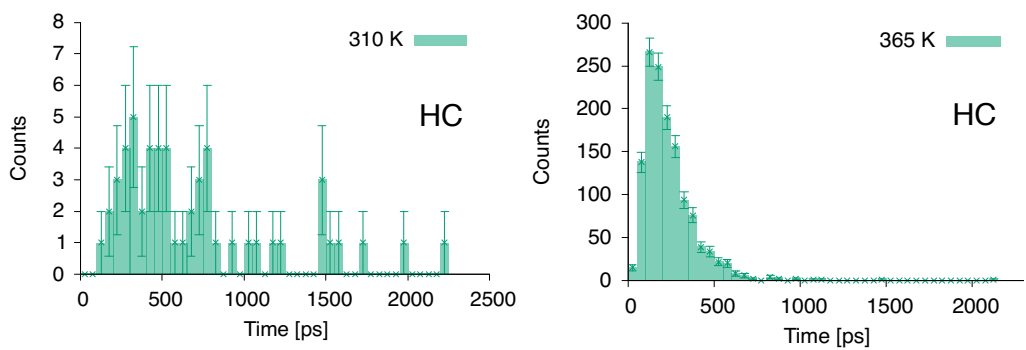

**Figure S4** – The distribution of travel times for the permeation events registered in the HC configuration, at 310 K (left) and 365 K (right)

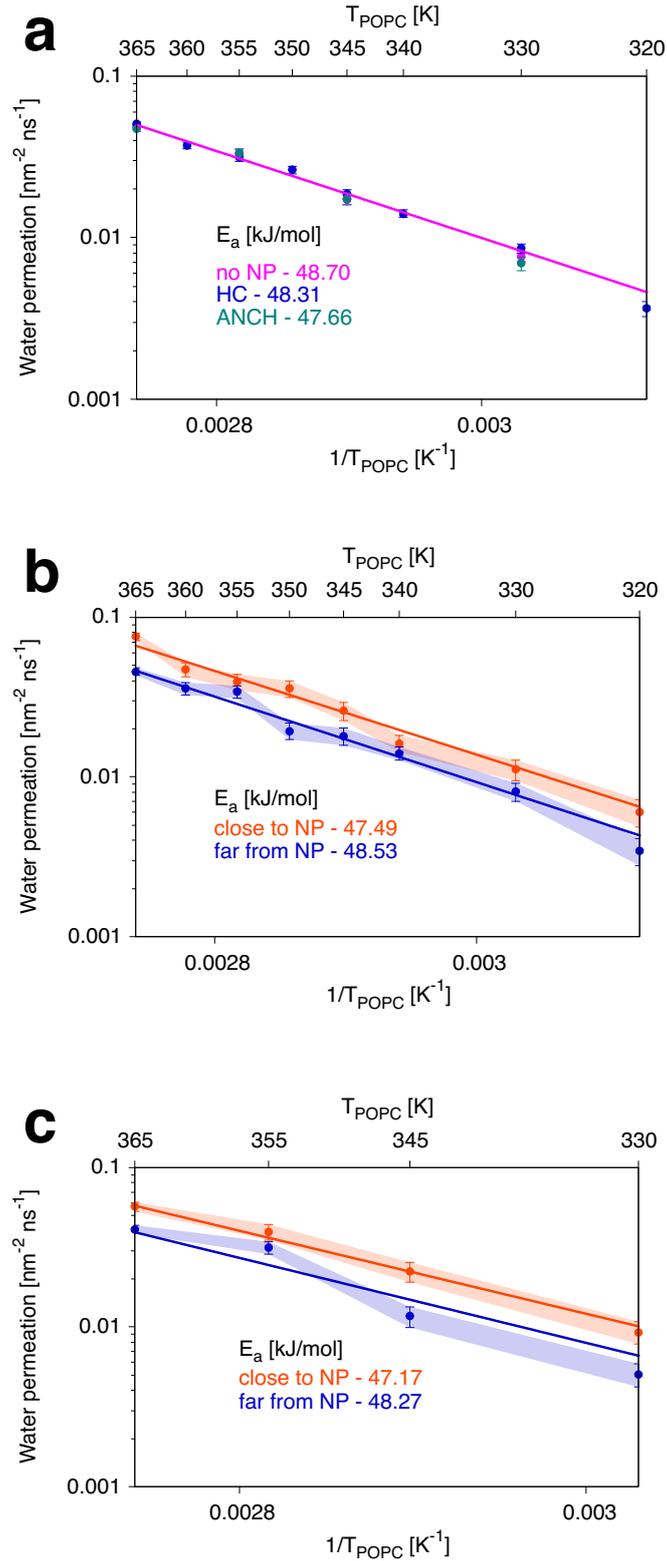

**Figure S5** – Arrhenius plot of the water permeation events. a. A comparison between the two NP-membrane configurations and the membrane with no NP; b. The comparison between the permeation events taking place close to the NP ($0 < d_r < 4$ nm) and far from it ($5 < d_r < 7$ nm), in the HC case; c. in the ANCH case.

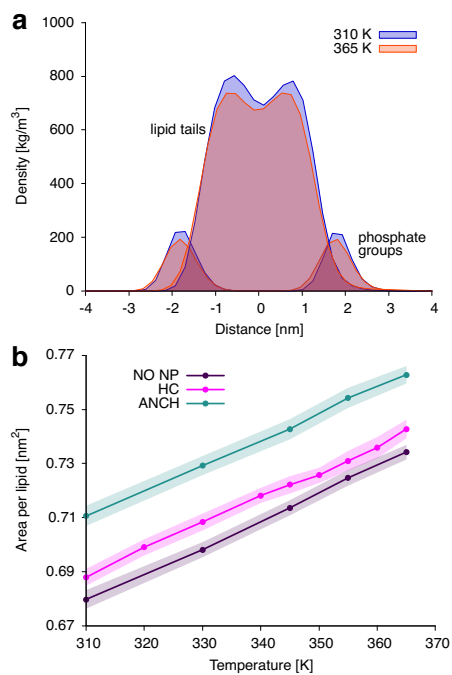

**Figure S6** – The increase of the average membrane temperature implies, as expected, little variations of the membrane thickness (a) and an increase of the area per lipid that is linear with temperature (b), in agreement with the experimental observations[3]